\documentclass[aps,prl,twocolumn,showpacs,a4paper,superscriptaddress,showkeys]{revtex4}

\usepackage{dcolumn,amsmath,xspace}

\usepackage{graphicx}
\usepackage{dcolumn}
\usepackage{amsmath}

\usepackage{units}

\newcommand{\degC}{\ensuremath{\!^{\circ}C}}

\begin{document}
\title{The structural properties of the multi-layer graphene/4H-SiC$(000\bar{1})$ system as determined
 by Surface X-ray Diffraction\\}

\author{J. Hass}
\affiliation{The Georgia Institute of Technology, Atlanta, Georgia
30332-0430, USA\\}

\author{R. Feng}
\affiliation{The Georgia Institute of Technology, Atlanta, Georgia
30332-0430, USA\\}

\author{J.E. Mill\'{a}n-Otoya}
\affiliation{The Georgia Institute of Technology, Atlanta, Georgia
30332-0430, USA\\}

\author{X. Li}
\affiliation{The Georgia Institute of Technology, Atlanta, Georgia
30332-0430, USA\\}

\author{M. Sprinkle}
\affiliation{The Georgia Institute of Technology, Atlanta, Georgia
30332-0430, USA\\}

\author{P. N. First}
\affiliation{The Georgia Institute of Technology, Atlanta, Georgia
30332-0430, USA\\}

\author{C. Berger}
\affiliation{The Georgia Institute of Technology, Atlanta, Georgia
30332-0430, USA\\}\affiliation{Institut N\'{e}el, BP166, 38042
Grenoble Cedex, France\\}

\author{W. A. de Heer}

\author{E. H. Conrad}
\affiliation{The Georgia Institute of Technology, Atlanta, Georgia
30332-0430, USA\\}

\begin{abstract}
We present a structural analysis of the multi-layer
graphene-4HSiC$(000\bar{1})$ system using Surface X-Ray
Reflectivity. We show for the first time that graphene films grown
on the C-terminated $(000\bar{1})$ surface have a
graphene-substrate bond length that is very short
($1.62\text{\AA}$). The measured distance rules out a weak Van der
Waals interaction to the substrate and instead indicates a strong
bond between the first graphene layer and the bulk as predicted by
{\it ab-initio} calculations.   The measurements also indicate
that multi-layer graphene grows in a near turbostratic mode on
this surface. This result may explain the lack of a broken
graphene symmetry inferred from conduction measurements on this
system [C. Berger et al., {\it Science} {\bf 312}, 1191 (2006)].
\end{abstract}
\vspace*{4ex}

\pacs{61.10.Kw, 68.55.-a, 68.35.-p, 61.46.-w, 61.10.Nz}
\keywords{Graphene, Graphite, X-ray reflectivity, X-ray
  diffraction, SiC, Silicon carbide, Graphite thin film}
 \maketitle
\newpage

\section{Introduction}
Recent experiments have demonstrated the unique electronic
properties of graphene
sheets.\cite{Berger04,Novoselov04a,Novoselov05b,Zhang05c,Berger06}
These works point to a potential route to a new nanoelectronics
paradigm based on an epitaxial graphene (EG).\cite{Berger04}  For
the purpose of this paper we define graphene as a single honeycomb
layer of graphite regardless of stacking order. At the moment
graphene is prepared either by mechanical exfoliation of flakes
from a bulk graphite sample that are subsequently deposited on an
insulating substrate\cite{Novoselov04a,Novoselov05b,Zhang05c} or
by sublimating Si from either of the polar faces of SiC; a process
that leaves a small number of graphene layers on the SiC
surface.\cite{Berger04} In the latter system, transport
measurements infer that the measured high mobilities are limited
to just a few graphene layers (perhaps only one), that must lie
near the SiC substrate. While there are similarities between the
magnetotransport properties of exfoliated graphene and SiC-grown
multi-layer graphene films, there are significant
differences.\cite{Berger06} For instance graphene layers grown on
different polar faces of SiC have electron mobilities that differ
by an order of magnitude.\cite{Hass06} Such graphene/substrate
specific transport properties strongly suggest that the substrate
interaction influences the electronic properties of the graphene
sheet. While this simple assertion may seem obvious, the structure
and influence of the interface remain points of heated conjecture.
One can ask if either the exfoliated or the SiC-grown multi-layer
graphene (or both) are really electronically the same as a ideally
isolated graphene sheet.  In other words, how does the interface
in both systems influence their electronic properties? In spite of
this debate there has been no direct structural characterization
of either the graphene-substrate interface or the graphene layers
themselves in either system. In this work we begin to address this
problem by performing a detailed investigation of the interface
structure of multi-layer graphene grown on the
4H-SiC$(000\bar{1})$ surface using Surface X-ray Diffraction
(SXD).

 Prior investigations of 6H- and
4H-SiC(0001) and $(000\bar{1})$ surfaces showed that multi-layer
graphene films can be grown on these surfaces by sublimating Si
from SiC during heating above $\sim\!\unit[1200]{\degC}$ in
ultrahigh vacuum (UHV).\cite{vanBommel75,Forbeaux00,Charrier02}
These studies show that multi-layer graphene grows epitaxially on
the (0001) Si-terminated (Si-face) surface of SiC, while
multi-layer graphene grown on the C-terminated $(000\bar{1})$
(C-face) surface is rotationally disordered and under some
conditions form nanocaps instead of a smooth
film.\cite{Kusunoki00} An explanation for the structural
differences for films grown on the two different faces was
proposed by Forbeaux et al.\cite{Forbeaux00}  Their conjecture is
that C-face multi-layer graphene becomes polycrystalline because
they have a stronger substrate-film bond compared to Si-face
graphite. The relative bond strengths were inferred from
K-Resolved Inverse Photoemission Spectroscopy
(KRIPES),\cite{Forbeaux00} and High Resolution Electron Energy
Loss Spectroscopy (HREELS) measurements.\cite{Angot_02} However,
recent work has shown that the C-face multi-layer graphene is not
polycrystalline and can be grown with domain sizes much larger
than those grown on the Si-face.\cite{Hass06} The improved
structural order of C-face films correlates with magnetotransport
measurements that to date find an order of magnitude improvement
in electron mobilities for films grown on the C-face compared to
Si-face films.\cite{Hass06}  Also, electronic coherence lengths
exceeding $\unit[1]{\mu m}$ have been measured for multi-layer
graphene films prepared on the C-face of SiC.\cite{Berger06} The
question becomes: how can a strongly bonded C-face film seemingly
ignore any substrate registry potentials and give rise to large
free rotating films?

Besides the question of topography differences, there are more
fundamental questions related to electron transport in these
films. For instance, how does charge transfer from the substrate
contribute to the doping of graphene near the surface? One of the
most important questions, and possibly related to the charge
transport question, is why transport measurements on multi-layer
graphene films grown on the $(000\bar{1})$ C-face seem to be
confined to just a few layers? Perhaps the most important question
is why conduction measurements suggest the existence of a Berry's
phase in three or more graphene layers.\cite{Berger06}  Similarly,
transport measurement infer\cite{Berger06}, and Angle Resolved
Photoemission measurements (ARPES) confirm\cite{Rollings_05}, the
existence of a Dirac Cone in the band structure of multi-layer
graphene films. These effects are not expected to occur in
multi-layer graphene system with bulk graphite {\it AB..}
stacking.\cite{McCann_PRL_06,Latil_PRL_06} In other words,
experiments suggests that the major conductor is either an
isolated single graphene sheet or possible {\it AA..} stacked
graphene layers.

In this paper we begin to address these questions.  We have
performed x-ray reflectivity experiments on the structure of
multi-layer graphene grown on the 4H-SiC$(000\bar{1})$ surface. We
find that the first layer of carbon with an areal density of
graphene sits very close to the last bulk SiC layer. For the
C-face the graphene-bulk spacing is found to be $1.62\pm
0.08\text{\AA}$. This number is consistent with recent {\it
ab-initio} calculations that also indicate a covalently bonded
first layer that is insulating and has no graphitic electronic
character.\cite{Varchon_PRL_07} We also demonstrate that the
C-face graphene films are flat with little or no corrugation in
contrast with mechanically exfoliated
graphite.\cite{Morozov_PRL_06} Also, by analyzing the graphite
inter-layer spacing, we can deduce that the graphene sheets are
stacked in a way resembling turbostratic graphite.

These results show that films grown on the C-face of SiC have a
strongly bonded very flat "buffer" layer. Subsequent graphene
layers can be rotationally disordered because of the weak registry
forces to this buffer layer.  Thus the strong bonding and
rotational disorder observed can be reconciled in a simple
structural model. Most important, the rotational disorder and
turbostratic character of the graphite suggest that the {\it AB..}
symmetry of the graphite is broken leaving a graphene character to
the films that may help to explain their conduction properties.

\section{Experimental}
All substrates were 4H-SiC purchased from Cree, Inc.\cite{Cree}
Prior to graphitization the $3\times 4\times 0.5$mm samples were
ultrasonically cleaned in acetone and ethanol.  Some samples were
hydrogen-etched prior to graphitization while others were not. The
etching removed all surface scratches and left a regularly stepped
surface but the graphite quality between etched and non-etched
samples is difficult to distinguish.\cite{Hass06}  Due to the
difficulty of preparing C-face 4H-SiC samples in UHV, they were
prepared by heating to \unit[1430]{\degC} in a vacuum RF-induction
furnace ($P = \unit[3\times10^{-5}]{\text{Torr}}$) for
\unit[5--8]{min}.\cite{Hass06} These parameters produced graphitic
films of 4--13 graphene layers. The thicker films grown on the
C-face reflects the current difficulty in producing less than 6
graphene layers in a furnace grown environment.\cite{Hass06}
Regardless of the film thickness, the interface and multi-layer
graphene film parameters measured were consistent as discussed in
the next sections. Once samples have been graphitized, they remain
inert allowing them to be transported into the Ultra High Vacuum
(UHV) chamber. The x-ray scattering experiments were performed at
the Advanced Photon Source, Argonne National Laboratory, on the
6IDC-$\mu$CAT UHV ($P < \unit[2\times10^{-10}]{\text{Torr}}$) beam
line at $16.2~$keV photon energy.

For all samples the number of graphene layers present was
determined by measuring the x-ray intensity as a function of
$\ell$ along the graphite $(1,\bar{1},\ell)_{G}$
rod.\cite{Charrier02} The notation $(h,k,\ell)_G$ identifies a
reciprocal-space point in reciprocal units ({\it r.l.u.}) of the
standard graphite hexagonal reciprocal lattice basis vectors ${\bf
q}=(h{\bf a}^*_G,k{\bf b}^*_G,\ell{\bf c}^*_G$, where
$a^*_G=b^*_G=2\pi/(a_G\sqrt{3}/2)$ and $c^*_G=2\pi/c_G$). The
nominal lattice constants for graphite are
$a_{\text{G}}=2.4589\text{\AA}$,
$c_{\text{G}}=6.674\text{\AA}$.\cite{Baskin_PR_55} For
reflectivity data we use unsubscripted reciprocal-space
coordinates $(h,k,\ell)$ that refer to the standard substrate
4H-SiC hexagonal reciprocal lattice units that are rotated
$30^\circ$ from the graphite reciprocal lattice basis. The
measured lattice constants were: $a_{\text{SiC}}=3.0802\pm
0.0006\text{\AA}$, $c_{\text{SiC}} = 10.081\pm .002\text{\AA}$ for
doped samples and $a_{\text{SiC}}=3.0791\pm 0.0006\text{\AA}$,
$c_{\text{SiC}} = 10.081\pm .002\text{\AA}$ for un-doped samples.
These are within error bars of published
values.\cite{Bauer_PRB_98}

\section{Results}
\begin{figure}[htbp]
\begin{center}
\includegraphics[width=8.5cm,clip]{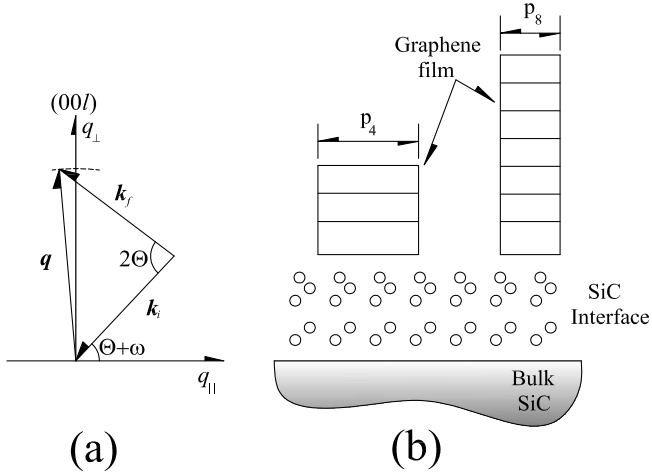}
\end{center}
\caption{(a) Schematic drawing of the reflectivity geometry.
Incident wave ${\bf k_i}$ strikes sample surface at an angle
$\Theta + \omega$. The diffracted wave, ${\bf k_f}$, is kept fixed
at $2\Theta$ from ${\bf k_i}$. ${\bf q}$ is "rocked" through the
$(00\ell)$ rod by rotating the sample through an angle $\pm
\omega$.  (b) Model of multi-layer graphene islands grown on a SiC
substrate with a reconstructed SiC interface layer. For specular
reflectivity, all $n$-layers islands can be represented as one
island with a fractional surface coverage parameter, $p_n$.}
\label{F:model}
\end{figure}

To obtain detailed information about both the graphene films and
the SiC-graphene interface, we have measured the surface x-ray
specular reflectivity from graphitized 4H-SiC$(000\bar{1})$.
Specular reflectivity only depends on the momentum transfer
perpendicular to the surface. The data is collected by integrating
rocking curves [see Fig.~\ref{F:model}(a)] around $q_\parallel = 0
$ for different perpendicular moment transfer vectors,
$q_\perp=2\pi\ell/c_{SiC}$, where ${\bf q}={\bf k}_f-{\bf k}_i$.
Since the reflectivity only depends on $q_\perp$, the data can be
analyzed using a one-dimensional model where all lateral
information is averaged over the $0.4\times 0.4$mm x-ray beam. The
scattered X-ray intensity $I(\Theta,q_\perp)$ is then,
\begin{equation}
I(\Theta,q_\perp)=A(\Theta,q_\perp)|F(q_\perp)|^2,
\label{E:intenisty}
\end{equation}
where $F(q_\perp)=F(\ell)$ is the total scattering amplitude from
the 4H-SiC substrate and the graphene film over layer.
$A(\Theta,q_\perp)$ is a term that contains all corrections due to
the experimental geometry. These include the Lorentz factor,
footprint correction, effective sample area, polarization factor,
and compensation for internal momentum transfer using critical
wave vectors for both the substrate and the
film.\cite{Robinson_review,Feng_thesis} $F(\ell)$ is a coherent
sum of the contributions from the graphene film and SiC substrate,
\begin{equation}
F(\ell)=e^{-2\gamma_{\text{SiC}}\sin^2{\pi \ell/m}}\{F_{SiC}(\ell)
+ \frac{\rho_{\text{G}}}{\rho_{\text{SiC}}}F_{G}(\ell)\}.
\label{E:ampl_1}
\end{equation}
$\rho_{\text{SiC}}$ and $\rho_{\text{G}}$ are the areal density of
a 4H-SiC$(000\bar{1})$ and a graphene (0001) plane, respectively.
The weighting factor, $\rho_{\text{G}}/\rho_{\text{Si}}$=3.132,
properly normalizes the scattered amplitude from the graphene
layer per 4H-SiC$(000\bar{1})$ $1\!\times\!1$ unit cell. The
exponential term in Eq.~(\ref{E:ampl_1}) is a roughness term that
assumes $c_{SiC}/m$ step fluctuations in the SiC surface
($\gamma_{\text{SiC}}$ is approximately the probability of finding
a step after traversing one SiC surface unit
cell).\cite{Feng_thesis} The step height is measured independently
from measurements of the specular rod full width at half-maximum
(FWHM) as a function of $\ell$ and gives the primary step height
to be $c_{SiC}/2$.\cite{conrad_review,Hass06}

To calculate $F_{SiC}(\ell)$ and $F_{G}(\ell)$ we use a model that
has a SiC substrate reconstruction layer and allows for patches of
the surface to be covered with different numbers of graphene
layers. The schematic model of the graphene covered SiC is shown
in Fig.~\ref{F:model}(b). In the model the SiC substrate
contribution is broken into two terms: (i) the amplitude from a
bulk terminated surface and (ii) the amplitude from a
reconstructed SiC interface layer.
\begin{equation}
F_{SiC}(\ell)=\frac{F_{SiC}^{bulk}(\ell)}{1-e^{-2\pi{i\ell}}}
+F_{SiC}^{I}(\ell).\label{beg2}
\end{equation}
The first term in Eq.~(\ref{beg2}) is the bulk 4H-SiC structure
factor, $F_{SiC}^{bulk}(\ell)$,\cite{Bauer_ACryst_01} modified by
the crystal truncation term,
$(1-e^{-2\pi{i\ell}})^{-1}$,\cite{Robinson_CTR_86} . The second
term in Eq.~(\ref{beg2}), $F_{SiC}^{I}(\ell)$, is the structure
factor of the reconstructed SiC surface. The SiC$(000\bar{1})$
face is known to reconstruct into a $2\!\times\!2$ structure near
the graphitization temperature.\cite{Forbeaux_SS_99} However, the
details of the reconstruction, and the nature of the SiC-graphene
interface are not known. Although we cannot obtain lateral
information about the interface structure from reflectivity data,
the vertical shifts of atoms and layer density changes associated
with them can be derived. To begin to understand this interface,
we allow for a reconstruction by placing two SiC bilayers between
the bulk and the multi-layer graphene film
[Fig.~\ref{F:model}(b)]. We then write the interface structure
factor as:
\begin{equation}
F_{SiC}^{I}(q_\perp)=
\sum_{j=1}^5{f_j\rho_je^{-c_j^2q_\perp^2}e^{iq_\perp z_j}},
\label{Surf_form}
\end{equation}
where $\rho_j$ is the relative atom density for the
$j^{\text{th}}$ interface layer ($\rho_j=1$ for a bulk layer) at a
vertical position $z_j$.$f_j$ is the atomic form factor of C or
Si. To allow for a reconstruction within each layer the term
$\text{exp}[-c_j^2q_\perp ^2]$ is added.  The $c_j$'s are the rms
vertical displacements of atoms due to a reconstruction in that
layer.\cite{robison 7x7}  The $5^\text{th}$ layer is added to
explore the possibility of adatoms between the SiC and the
graphene.

\begin{figure}[htbp]
\begin{center}
\includegraphics[width=8.0cm,clip]{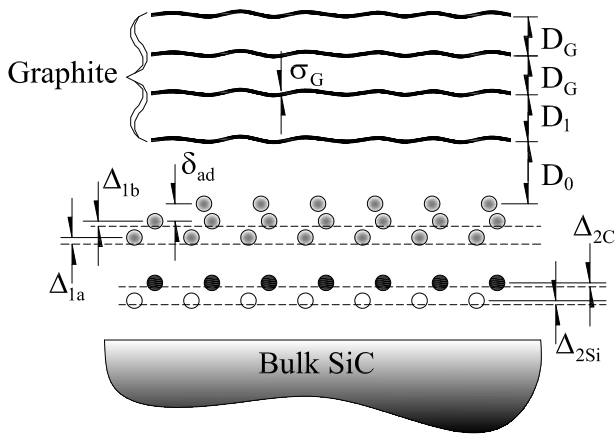}
\end{center}
\caption{A schematic model of multi-layer graphene grown on the
4H-SiC$(000\bar{1})$ substrate. Dashed lines are the bulk SiC
lattice planes before interface relaxation ($\Delta$'s). The
$5^\text{th}$ (adatom) layer is displaced $\delta_{ad}$ from the
last interface layer. ($\bullet$) are carbon atoms and ($\circ$)
are silicon atoms. The shaded circles in the top three layers of
the interface can be either carbon or silicon atoms depending on
which of the two models, Carbon-corrugated or Carbon-rich, is
used.} \label{F:C-model_sch}
\end{figure}

To be completely general, we allow the scattered amplitude,
$F_{G}(\ell)$, from the graphene film in Eq.~(\ref{E:ampl_1}) to
take into account the possibility of a spatial distribution of
varying graphene layers. This is done by defining an occupancy
parameter $p_n$ as the fractional surface area covered by all
graphene islands that are $n$ graphene layers thick. $p_n$ is
subject to the constraint equation $\sum{p_n}=1$, where $p_0$ is
the fraction of area that has no graphene. The multi-layer
graphene structure factor can then be written in the general form:
\begin{subequations}
\label{E:island_I}
\begin{equation} F_{G}(\ell)=f_{C}\left(
\sum_{n=1}^{N_{\text{max}}}p_{n}e^{-q_\perp^2\sigma_{G}^{2}/2}
\sum_{m=1}^{n}e^{2\pi{il}z_m/c}\right), \label{E:Fpb}
\end{equation}
\begin{equation}
z_m=\left\{\begin{array}{ll}
D_0 + (m-1)D_1 &\mbox{$m \leq 2$} \\
D_0 + D_1+(m-2)D_G &\mbox{$m > 2$}\label{E:beg4}
\end{array}
\right. .
\end{equation}
\label{E:grap_stru}
\end{subequations}
$f_{C}$ is the atomic form factor for carbon and $N_{\text{max}}$
is the number of graphene layers in the highest island. The
coordinates $z_m$ in Eq.~(\ref{E:grap_stru}) are the position of
the $m^{\text{th}}$ graphene atomic layer relative to the last
plane of SiC interface atoms. $D_0$ is the spacing between the
bottom layer of an island and the substrate and $D_G$ is the
average layer spacing between graphene layers in an island (see
Fig.~\ref{F:C-model_sch}). We have allowed the spacing between the
first and second layer graphene, $D_1$, to be different from the
bulk to allow for differences due to a different bonding geometry
with the substrate. Because STM studies of multi-layer graphene
films grown on the Si-face indicate some buckling of the graphite
layer,\cite{STM_buckeling} we also allow for a small vertical
height distribution in each graphene layer. This is modelled
similar to the interface relaxations in Eq.~(\ref{Surf_form}) by a
Debye-Waller term, $\sigma_G$.  As we will show, C-face graphene
films show no significant buckling.

\begin{figure}
\begin{center}
\includegraphics[width=8.0cm,clip]{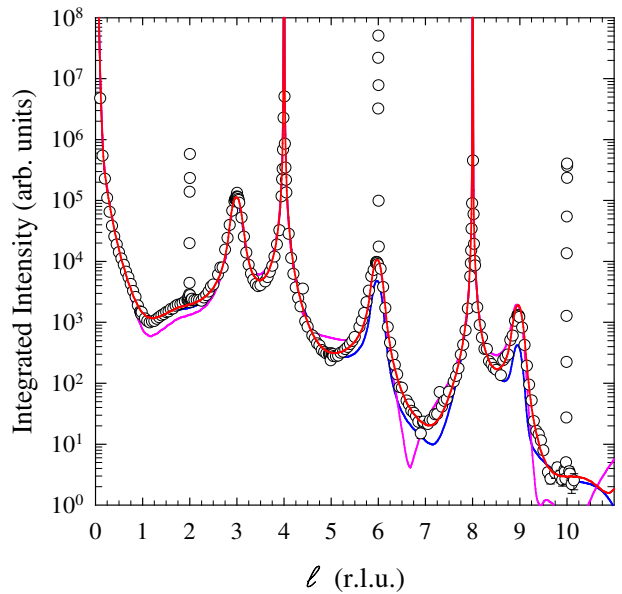}
\end{center}
\caption{Specular Reflectivity vs. $q_\perp$ (in r.l.u.) for
graphitized 4H-SiC$(000\bar{1})$ C-face surface with 9 graphene
layers.  Solid lines are best fits to the structural models
described in the text. Red solid line is the best fit to the
structural model with a smooth graphene layer
($\sigma_G=0.0\text{\AA}$). Blue solid line is the best fits with
a corrugated graphite layer ($\sigma_G=0.25\text{\AA}$).  Magenta
line is the best fit if the graphene substrate distance $D_0$ is
reduced $10\%$.} \label{F:reflec_1}
\end{figure}

Reflectivity data for a C-face multi-layer graphene film are shown
in Fig.~\ref{F:reflec_1}. The main bulk 4H-SiC peaks occur at
$\ell = 4$ and $\ell = 8$.  The sharp peaks at $\ell = 2$, 6 and
10 are the "quasi-forbidden" reflections of bulk
SiC.\cite{Bauer_ACryst_01} The graphite bulk reflections are
expected at $\ell \sim 3$, 6 and 9. While there are many variables
in Eqs.~(\ref{E:intenisty})-(\ref{E:grap_stru}) that eventually
must be fit, a number of the parameters are quite unique and
insensitive to the exact structural model used for the
SiC-graphene interface. For instance, because the graphite Bragg
points are intense and narrow in $\ell$, the mean spacing between
graphene layers, $D_G$, is determined with high accuracy relative
to the known SiC lattice constant. Similarly, the graphene layer
roughness or corrugation, $\sigma_G$, is determined almost solely
by the intensity decay of the graphite Bragg points as a function
of $\ell$. Once these nearly model-independent parameters are
determined, they are fixed so that different structural models of
the interface can be compared without relying on adjusting large
numbers of parameters.

We have tested a number of structural models for the
graphene/4H-SiC$(000\bar{1})$ interface.  Simple relaxations of
the top SiC bi-layers always give poor fits to the data. Even
attempts to make a carbon rich phase that extends many layers into
the bulk, a model that has been proposed in the
literature\cite{vanBommel75,Forbeaux98,Barrett_PRB_05}, was not
compatible with the data. The best fit model is a distorted
bilayer between the graphene and bulk SiC.  A schematic of the
model is shown in Fig.~\ref{F:C-model_sch}. In this model the
first bilayer above the bulk is slightly relaxed. However, the
next bilayer (immediately below the graphene) has a significant
relaxation. As we will show below, two similar versions of this
model structure give nearly identical fits to the data.

Before looking at the details of these models, we point out a few
important model-independent parameters for the graphite film.
First, the average graphene inter-layer spacing is found to be
$D_G=3.368 \pm .005 \text{\AA}$. This and other graphite film
parameters are given in Table~\ref{tab:C_graphite}.  The value was
determined from samples with films ranging from 9-13 graphene
layers (averaged over the beam footprint). As mentioned above the
inter-layer spacing is nearly independent of all other fit
parameters and can be determined with high accuracy because it is
fixed by the $\ell$ position of the three strong graphite Bragg
peaks in Fig.~\ref{F:reflec_1}. The inter-layer spacing is larger
than bulk crystalline graphite
($3.354\text{\AA}$)\cite{Baskin_PR_55} but smaller than the
lattice spacing of turbostratic graphite
($D_{TG}=3.440\text{\AA}$).\cite{Franklin_51,Cancado_PRB_02} The
larger spacing is due to stacking faults between adjacent layers
caused by interference between $\pi^*$ states that give rise to a
larger repulsive interaction between adjacent graphene
sheets.\cite{Franklin_51}

\begin{table}
\caption{\label{tab:C_graphite}Structural parameters for graphene
grown on 4H-SiC$(000\bar{1})$ C-Face.  Parameters are defined in
Fig.~\ref{F:C-model_sch}}
\begin{ruledtabular}
\begin{tabular}{cccccc}
 &$D_0$ (\AA)&$D_1$ (\AA)&$D_G$(\AA)&$\sigma_{G}$ (\AA)&\\
\hline fit value& 1.62 & 3.41&3.368 & 0.00  &\\
uncertainty& 0.08 & 0.04  & 0.005& 0.05   &\\
\end{tabular}
\end{ruledtabular}
\end{table}

Another parameter that is insensitive to the details of the model
is, $\sigma_G$, in Eq.~(\ref{E:Fpb}). This parameter can be
interpreted two ways: either as a finite width of a graphene layer
due to buckling of carbon atoms in the layer, or as an RMS
roughness of a graphene layer due to vertical disorder over the
coherence length of the x-ray beam ($\sim 2\mu\text{m}$). We find
that $\sigma_G=0.0\pm 0.05\text{\AA}$ (see
Table~\ref{tab:C_graphite}). Because of the exponential form in
Eq.~(\ref{E:Fpb}), a finite layer width manifests itself as a
rapid decay in the graphite Bragg peak intensity at high $\ell$.
This is demonstrated in Fig.~\ref{F:reflec_1} where we compare a
flat graphite film to a film with an RMS thickness of
$\sigma_G=0.25\text{\AA}$.  The finite layer width severely
reduces the graphite peak intensities at $\ell =$6 and 9.

\begin{figure}[htbp]
\begin{center}
\includegraphics[width=7.5cm,clip]{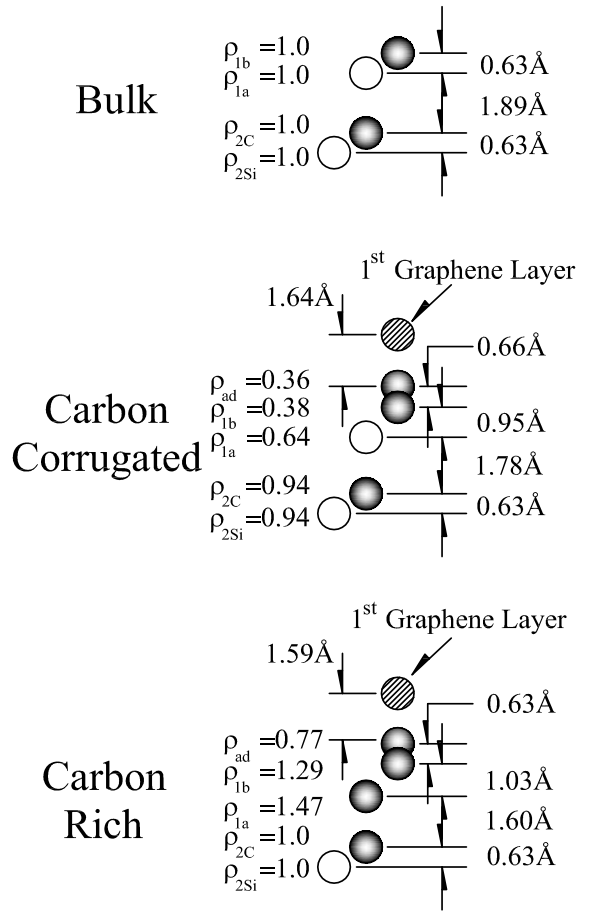}
\end{center}
\caption{Schematic ball models of bulk, C-Corrugated and C-Rich
interface layers between the substrate and the graphene film.
($\bullet$) are carbon atoms and ($\circ$) are silicon atoms.
Hatched atoms are carbon atoms in the first graphene layer.
Interlayer spacings and densities (relative to bulk SiC) are
shown.} \label{F:C-Ball-model_sch}
\end{figure}

Fits to the reflectivity show that two similar model structures
for the interface region between the bulk and the graphene
represent the experimental data equally well.  We refer to these
models as the "Carbon-corrugated" and "Carbon-rich" models.  In
both models the SiC bilayer immediately above the bulk in
Fig.~\ref{F:C-model_sch} remains "bulk-like" in terms of both
density and inter-layer spacing.  The two models are distinguished
by the structure of the next three layers just below the graphene
film. Ball models of the two structures are shown in
Fig.~\ref{F:C-Ball-model_sch} and the detailed fitting parameters
are given in Table \ref{tab:table1}.  Structural values were
determined for three different samples. The fitting parameter
variations from sample to sample are included in the uncertainty
limits of Table~\ref{tab:table1}.

\begin{table*}
\caption{\label{tab:table1}Best-fit interfacial structural
parameters for graphite covered 4H-SiC$(000\bar{1})$ (C-Face).
Data for both the "C-Corrugated" and "C-Rich" models give nearly
identical fits. All fits find $c_j\sim 0\text{\AA}$ for all
layers. Parameters are defined in Fig.~\ref{F:C-model_sch}}
\begin{ruledtabular}
\begin{tabular}{cccccccccc}
&$\delta_{ad}$ (\AA)&$\rho_{ad}$
(\AA)&$\Delta_{1b}$(\AA)&$\rho_{1b}$ (\AA)
&$\Delta_{1a}$ (\AA) &$\rho_{1a}$ &$\Delta_{2C}=\Delta_{2Si}$(\AA)& $\rho_{2Si}=\rho_{2C}$ &\\
\hline C-corrugation&  0.66 & 0.36&0.18 & 0.38 &-0.14
& 0.64 & -0.03 & 0.94 &\\
Atom Type&   & carbon& & carbon &
& silicon& &  &\\
\hline C-rich&  0.63 & 0.77&0.11 & 1.29 &-0.33
& 1.47 & -0.04 & 0.94 &\\
Atom Type&   & carbon& & carbon &
& carbon& &  &\\
\hline uncertainty& 0.04& 0.08 & 0.04  &0.08
& 0.04 & 0.10 & .04 & .05 &\\
\end{tabular}
\end{ruledtabular}
\end{table*}

\begin{figure}[htbp]
\begin{center}
\includegraphics[width=8.0cm,clip]{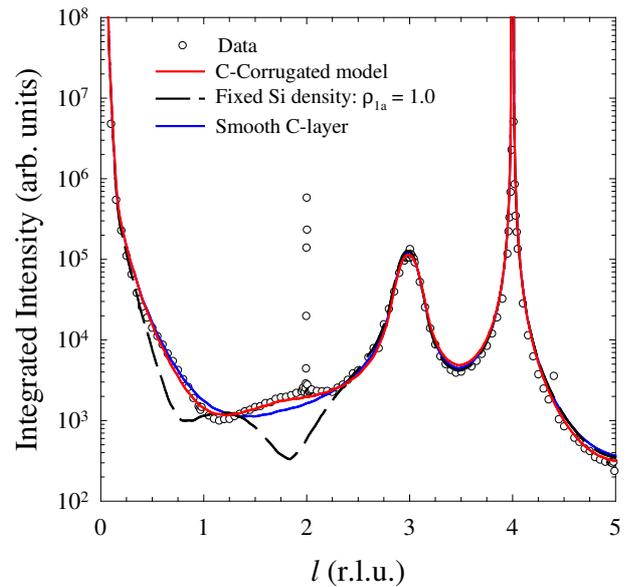}
\end{center}
\caption{Specular Reflectivity for graphitized
4H-SiC$(000\bar{1})$ C-face surface. ($\circ$) are the data. Red
line is the best fit to the Carbon-corrugated top layer. The black
dashed line show the fit for the same model if the Si layer
density is fixed at the bulk value ($\rho_{1a}=1$). The blue line
is a fit when the carbon corrugation in the top layer is removed
but the total density remains the same ("Smooth C-layer").}
\label{F:reflec_C-face_model}
\end{figure}

In the C-Corrugated model the last SiC bilayer is contracted
inwards towards the bulk by $0.11\text{\AA}$ to give a slightly
smaller Si--C bond length.  In the uppermost bilayer the carbon is
buckled into two equal density layers. The density of both the Si
layer ($\rho_{1a}$) and the sum of the buckled carbon layers
($\rho_{ad}+\rho_{1b}$) in this bilayer are each $\sim 2/3$ of the
bulk value.  It is unlikely that the last layer is a carbon
adatom.  If it were, the density required to saturate the dangling
bonds in the carbon layer below would be $\rho_{ad}=\rho_{1b}/3$
instead of being equal.  For this reason we refer to the model as
a corrugated surface. We note that the fits are very sensitive to
the Si density, $\rho_{1a}$, in the last bilayer. If we force the
last bilayer to have the same Si atom density as in the bulk, the
best fit model cannot reproduce the data. This is demonstrated in
Fig.~\ref{F:reflec_C-face_model} where we show a best fit to the
"C-Corrugated" model but force $\rho_{1a}$ to be the bulk
denisity. Similarly, removing the buckling in the carbon layer
("Smooth C-layer" model) while keeping the total density the same
cannot reproduce the reflectivity modulation between
$0.5<\ell<2.5$ (see Fig.~\ref{F:reflec_C-face_model}).

To first order the ratio of the atomic form factors for Si and C,
$f_{Si}/f_{C}$ in Eq.~(\ref{Surf_form}), are determined simply by
the ratio of their atomic numbers $14/6=2.33$.  Therefore, the
model calculation should give a similar fit if the Si atoms in the
top SiC bilayer are replaced by carbon atoms with 2.33 times the
density ($\sim 2.33\times 0.64=1.49$).  This replacement gives the
"C-rich" model shown in Fig.~\ref{F:C-Ball-model_sch} with
densities and layer spacings adjusted to give the best fit to the
data. In Fig.~\ref{F:C-Ball-model_sch} the best fit parameters
show that there are two main differences between the C-Corrugated
and C-Rich models.  First, the layer spacings between bi-layers is
considerably shorter ($1.60\text{\AA}$) and second, the densities
in the last layers are higher. The bilayer spacing measured in the
C-rich model is slightly larger than the bond length of diamond
($1.54\text{\AA}$).\cite{Burdett} The higher carbon layer
densities have a similar significance in that they lie half way
between the SiC density ($\rho=1.0$) and that of graphene
($\rho=3.13$).  In fact, the first C-layer in the bilayer has a
density close to the atom density of a (111) diamond plane, 1.51.

\begin{figure}[htbp]
\begin{center}
\includegraphics[width=8.0cm,clip]{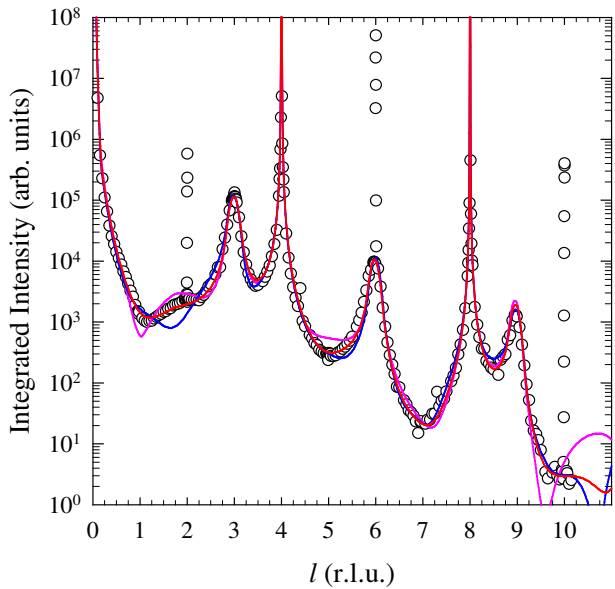}
\end{center}
\caption{A comparison of the calculated reflectivity vs. $q_\perp$
(in r.l.u.) for different first bilayer models.  Red line is the
best fit structure with bulk bilayer parameters.  Blue line is a
fit with $\rho_{2Si}$ fixed at a value $25\%$ less than the bulk.
Magenta line is a best fit with the both $\Delta_{2Si}$
and$\Delta_{2C}$ relaxed towards the bulk by $5\%$.}
\label{F:first_bi-layer}
\end{figure}

While it may seem reasonable to expect that as Si sublimates from
the surface a carbon rich interface forms with some diamond-like
character, we should caution that there are other ways to
interpret these results.  First of all, the spacing between planes
in the bilayer is much larger, 0.63-1.03\AA, while in diamond they
should be much lower, 0.51\AA. The C-Rich phase is also
considerably different from the "extended diamond phase" proposed
in the literature because it does not extend beyond the first
bilayer.\cite{vanBommel75,Forbeaux98,Barrett_PRB_05} In both
models the relaxation of the bilayer above the bulk is small,
contrary to what might be expected if there were significant
density changes in that layer.  These small changes from the model
are not due to an insensitivity to either the layer spacings or
the layer density.  This is shown in Fig.~\ref{F:first_bi-layer}
where we compare calculated best-fit reflectivities when either
the Si density $\rho_{2Si}$ is reduced or the Si--C spacing
$\Delta_{2Si}$ is changed from the ideal value.  As can be seen,
inter-planar spacing changes of less than $5\%$ ($<0.1\text{\AA}$)
lead to obviously poor fits. Similarly, reducing the Si atom
density in this layer by more than $25\%$ makes the fit much
worse.  Therefore, the interfacial layer does not extend much
beyond the top-most SiC bilayer. Note also that the total layer
density of the last three interface layers is
$\rho=1.47+1.29+0.77=3.53$. This density is slightly larger than
the density of a graphene sheet ($\rho=3.13$).  Rather than
thinking of this layer as an ideal diamond like layer, it may be
more appropriate to view it as a buckled graphene sheet with a
mixture of $sp^2$ and $sp^3$ bonded carbon.

The most important finding from this work is that the first
graphene layer sits above the last bulk carbon layer at a distance
of $D_0=1.62\pm 0.08\text{\AA}$.  This value is, within error
bars, insensitive to which structural model is used and can be
determined with reasonable sensitivity as demonstrated in
Fig.~\ref{F:reflec_1}. The figure shows that $10\%$ variations in
$D_0$ from its optimal value lead to very poor fits to the data.
The very short bond distance measured suggests that the first
graphene layer is not simply bonded to the substrate with Van der
Waal's bonds but instead has a much stronger interaction with the
substrate.

\section{Discussion}
\label{S:Disc} The x-ray reflectivity data shows that the
interface between epitaxial graphene and the 4H-SiC$(000\bar{1})$
substrate is sharp.  The interface is comprised of no more than
1--2 SiC bi-layers.  The graphene that grows is flat (i.e.
$\sigma_G=0\text{\AA}$) except for a small potential buckling of
the first layer. There are two key structural parameters that
deserve attention. The first is the inter-layer spacing between
graphene sheets that is much larger than expected for {\it AB...}
stacked graphene layers and points to a significant stacking fault
density in the film. Because stacking faults cause interference
between $\pi^*$ states in adjacent layers, these layers have a
larger spacing. The mean layer spacing can, therefore, be used to
estimate the stacking fault density.\cite{Franklin_51} If we
define the probability, $\gamma$, that any two adjacent sheets are
faulted, then the inter-layer spacing will range from that of {\it
AB...} stacked graphite ($3.354\text{\AA}$) when $\gamma=0$ to
that of turbostratic graphite ($3.44\text{\AA}$) when $\gamma=1$.
In that case the average inter-layer spacing for some finite
number of stacking faults is approximately;\cite{Franklin_51}
\begin{equation}
D_G=3.44-0.086(1-\gamma ^2).
\end{equation}
Using the measured $D_G=3.368\text{\AA}$, gives $\gamma=0.4$ for
these C-face films. In other words, after every $1/(1-\gamma) =
1.6$ graphene sheets, a stacking fault occurs in the film. The
fact that there are frequent stacking faults is not surprising
since there is significant rotational disorder of graphene layers
grown on this surface.\cite{Forbeaux00, Hass06} A pair of graphene
sheets that are rotated with respect to each other would lead to
regions of local {\it AB...} stacking separated by regions with
other stacking arrangements. The mean graphite inter-layer spacing
would then be determined by the degree of rotational disorder.
Experiments to quantify the stacking and rotational disorder are
currently underway.

The existence of a large stacking fault density has an important
bearing on the results of conductivity experiments on C-face grown
multi-layer graphene films.  Magnetotransport \cite{Berger06},
Infared Spectroscopy (IRS) \cite{Sadowski_PRL_06} and
photoemission experiments \cite{Rollings_05} indicate that
multi-layer graphene films grown in SiC behave like isolated
graphene sheets. In the photoemission experiments a clear Dirac
dispersion cone is measured.  The origin of this type of
dispersion in the electronic band structure is the symmetry of
carbon bonding in a single graphene sheet. {\it AB...} stacking in
multi-layer graphene films (i.e graphite stacking) would break
that symmetry and causes significant changes to the band
structure, even for a few layers.\cite{McCann_PRL_06,Latil_PRL_06}
In the multi-layer graphene films grown on the C-face of SiC, the
{\it AB...} stacking disorder may inhibit symmetry breaking and
allow sheets in the film to behave electronically as if they were
physically isolated.\cite{Latil_PRL_06}

In addition to the stacking fault density, the short bond length,
$D_0$, between the interface and the first graphene sheet
indicates an additional way the graphene sheets become isolated
from the substrate.  While the {\it AB...} stacking in bulk
graphite breaks the hexagonal symmetry of an isolated graphene
sheet, it has been assumed that the substrate-graphene interaction
will have a similar effect.\cite{Latil_PRL_06} Indeed the short
$D_0$ bond length measured here for the
graphene/4H-SiC$(000\bar{1})$ interface implies a strong
interaction that is consistent with Photoemission
results.\cite{Forbeaux00} The short bond length we measure for the
C-face films has been recently confirmed by {\it ab-initio}
calculations.\cite{Varchon_PRL_07} Those calculations show that
when a single graphene layer is grown on a SiC substrate the
material remains insulating.  The Dirac cone dispersion, of an
isolated graphene sheet does not appear until a second graphene
layer is formed.\cite{Varchon_PRL_07} Therefore, the first carbon
layer with a graphene density acts like a "buffer" layer between
the substrate and the rest of the graphene film. From the
structural properties of the graphene/SiC interface measured here,
a model emerges that may explain the graphene character of these
films seen in magnetotransport, IRS, and Photoemission
measurements as well as in band structure calculations. In this
model, one or two graphene layer, primarily responsible for the
conduction, lies between the "buffer" layer and the imperfectly
stacked graphene layers above it.

While the nature of the buffer layer is not completely
characterized, the reflectivity data offer two possibilities. (1)
In the C-Corrugated model the buffer layer is simply the first
flat graphene layer above the interface.  The SiC bilayer below
this layer is relaxed with a lower density of atoms compared to
the bulk.  (2) In the C-rich structure, a highly buckled carbon
layer, with a total carbon density nearly equal to graphene,
resides between the substrate and the rest of the film. Low Energy
Electron Diffraction (LEED) experiments show that UHV-grown
multi-layer graphene on the C-face surface exhibits a $2\times 2$
reconstruction.\cite{Forbeaux_SS_99}  Our x-ray measurements
confirm that a $2\times 2$ structure still persists even when the
films are thick enough that LEED is no longer capable of probing
the interface.\cite{Hass06}  The long range order of the $2\times
2$ reconstruction is $\sim 200\text{\AA}$.  This is much smaller
than the film structural coherence length of $\sim 3000\text{\AA}$
and suggests that the interface is never fully ordered.  A
possible explanation maybe that different parts of the surface are
in different stages of graphitization.

It is significant that the RMS layer width of the graphene is
zero, $\sigma_G$ in Eq.~(\ref{E:Fpb}).  $\sigma_G$ can represent
either a random film roughness or a RMS corrugation of the
graphene that is commensurate with the substrate.  Because it is
zero, we can conclude that beyond the buffer carbon layer the
graphene layers are flat and must be very weakly interacting with
any substrate potential.  This explains why C-face graphene films
can be rotationally disordered but have large domain sizes.  The
energy cost per atom to rotate a graphene sheet on a flat graphene
substrate is very low (
$<50$meV/atom).\cite{Girifalco_JCP_56,Kolmogorov_PRB_05}  At the
growth temperatures of 1400C, and given the low registry forces
implied from these experiments, growing graphene sheets can rotate
freely, rather than becoming polycrystalline as suggested by
Forbeaux et al.\cite{Forbeaux00} On Si-face multi-layer graphene
films the situation is different.  There is a $(6\sqrt{3} \times
6\sqrt{3})\text{R}30^0$ reconstruction in the first 2-3 graphene
layers on this surface.\cite{vanBommel75,Tsukamoto_97}  The
graphene has a nonzero corrugation of about
$0.25\text{\AA}$\cite{STM_buckeling} that could be enough to lock
the growing film into registry.  Step boundaries or other defects
in the substrate can put domain boundaries in the graphene that
are not easily removed by rotating large areas of the film.

\section{Summary}
In this work we have measured a number of important structural
parameters of multi-layer graphene grown on the carbon terminated
4H-SiC$(000\bar{1})$ surface. The most important finding is that
the separation between the first graphene layer and the SiC
surface is very short (1.62\AA ). This distance is not much larger
than a diamond bond length implying a very strong interaction
between the first graphene layer and the substrate.  It is
consistent with recent band structure calculations that show a
large energy gap for a single graphene layer on the
4H-SiC$(000\bar{1})$ surface.\cite{Varchon_PRL_07}  Subsequent
graphene layers have an RMS corrugation (averaged over 9 layers)
that is less than 0.05\AA.  This suggest that the strongly bound
buffer graphene layer shields subsequent layers from the interface
corrugation potential.  Therefore, unlike exfoliated graphene
flakes deposited on $\text{SiO}_2$\cite{Morozov_PRL_06}, the
graphene layers grown by sublimation on the C-face of SiC are very
smooth.

The graphene films show evidence of a large density of stacking
faults.  While the topography of these faults remains to be
determined, it does suggest that the {\it AB...} stacking of bulk
graphite is not present in these films.  This may be the reason
why Dirac electrons, expected only in isolated graphene sheets
where {\it AB...} stacking does not break the graphene symmetry,
are seen in this multi-graphene system.

\subsection*{Acknowledgments}
This research was supported by the National Science Foundation
under Grant No. 0404084, and by Intel Research.  The Advanced
Photon Source is supported by the DOE Office of Basic Energy
Sciences, contract W-31-109-Eng-38.  The $\mu$-CAT beam line is
supported through Ames Lab, operated for the US DOE by Iowa State
University under Contract No.W-7405-Eng-82.  Any opinions,
findings, and conclusions or recommendations expressed herein are
those of the authors and do not necessarily reflect the views of
the research sponsors.

\end{document}